\RequirePackage{lineno}
\documentclass[aps,prb,twocolumn,showpacs,superscriptaddress]{revtex4}

\usepackage[english]{babel}
\usepackage[utf8]{inputenc}
\usepackage[T1]{fontenc}
\usepackage{amsmath, amsthm, amssymb}
\usepackage{graphicx}
\usepackage{dcolumn}
\usepackage[usenames,dvipsnames]{color}
\usepackage{bbm}
\usepackage{url}
\usepackage{float}
\usepackage{xcolor}
\usepackage{tabularx}
\usepackage{upgreek}

\usepackage[version=3]{mhchem}

\newcommand{\kett}[1]{\ensuremath{\left| #1 \right.\rangle}}				                  % Ket - Vektor --> for arrows
\newcommand{\ket}[1]{\ensuremath{\vert \,#1\, \rangle}}				                        % Ket - Vektor

\begin{document}
%\title{Detailed Analysis of Critical Points in Coupled 2D Spin Dimer Systems and the Occurrence of BKT Transition}
\title{Magneto-caloric effects, quantum critical points, and the 
Berezinsky-Kosterlitz-Thouless transition in 2D coupled spin dimer systems}
\author{Dominik Stra\ss el}
\email{strassel@physik.uni-kl.de}
\affiliation{Department of Physics and Research Center Optimas, Technical University Kaiserslautern, 67663 Kaiserslautern, Germany}
\author{Peter Kopietz}
\affiliation{Department of Theoretical Physics, Goethe-University Frankfurt, 60438 Frankfurt, Germany}
%\affiliation{Department of Physics, University of Florida, 32611 Gainesville Florida, USA}
\author{Sebastian Eggert}
\affiliation{Department of Physics and Research Center Optimas, Technical University Kaiserslautern, 67663 Kaiserslautern, Germany}

\date{\today}

\begin{abstract}
Spin dimer systems are a promising playground for the detailed study of 
quantum phase transitions. 
Using the magnetic field as the tuning parameter
it is in principle possible to observe a crossover from the characteristic scaling 
near critical points to the behavior of a finite temperature phase transition. 
In this work we study two-dimensional coupled spin dimer systems
by comparing numerical quantum 
Monte Carlo simulations with analytical calculations of the 
susceptibility, the magneto-caloric effect, and 
the helicity modulus.  The magneto-caloric behavior of the magnetization with temperature
can be used to determine the critical fields with high accuracy, but the critical 
scaling does not show the expected logarithmic corrections.  The zeros of the cooling rate
are an excellent indicator of the competition between quantum criticality and
vortex physics, but they are not directly associated with the quantum phase transition or
the finite temperature Berezinsky-Kosterlitz-Thouless transition.
The results give a unified picture of the full quantum and finite temperature phase diagram.
\end{abstract}

\pacs{75.10.Jm, 75.30.Sg, 75.30.Kz, 05.30.Jp}
\maketitle

\section{Introduction}\label{sec:intro}
The study of quantum phase transitions (QPT) remains a very active topic in 
many fields of physics,
spurred by experimental progress to create novel tunable interacting 
systems.  
QPT occur in quite different materials,
including heavy fermion compounds,\cite{} unconventional superconductors,\cite{} 
Mott insulators,\cite{}
coupled spin systems,\cite{} and ultracold atoms.\cite{} 
In particular, the common phenomenon of Bose Einstein condensation (BEC) of
strongly interacting bosons by tuning the interaction or the chemical potential 
can now be found in a range of 
different physical systems.  
Ultracold atomic gases allow the tuning of 
interactions via Feshbach resonances,\cite{} but also cross-dimensional phase
transitions \cite{ott} and
Berezinsky-Kosterlitz-Thouless (BKT) behavior \cite{dalibard}
have been observed recently. 
%The BEC of magnons give new insights into the dynamic behavior \cite{}.
 Phase transitions in coupled spin dimer systems are prime examples of 
BEC of strongly interacting triplons,\cite{Ruegg03p62,lang,sachdev,PhysRevLett.87.206407,Haseda1969,Tachiki1970} 
which allow easy 
tuning of the chemical potential via the magnetic field.
Although QPT's occur at zero temperature as a function of 
a non-thermal control parameter such as
the interaction, 
effective mass, or the chemical potential, a characteristic critical scaling
with temperature can be observed in a large range above the critical point.\cite{sachdev}
In general a detailed analysis is necessary in order to understand how the 
critical behavior is reflected in the experiments and if the finite-temperature
phase transition is affected in the vicinity the QPT, where thermal fluctuations
are comparable to quantum fluctuations. Compared to bosonic gases of atoms and magnons
the temperature control is relatively easy in triplon gases, which allows a systematic
analysis of the critical scaling behavior near the QPT.

In this paper we focus on the theoretical analysis of quantum critical points of 
antiferromagnetic spin dimer systems which are weakly 
coupled in two-dimensions.  Two QPT's
can be observed:  As the field is increased through the lower critical value $B_c$ 
the spin dimers start to be occupied by 
triplons and the magnetization increases with characteristic two-dimensional
logarithmic behavior.  The second QPT corresponds to the saturation field $B_s$.
The intermediate phase is characterized by long-range phase coherence of triplons at $T=0$
and BKT behavior\cite{Bere1,Bere2,KT1,KT2} at finite $T$.
Similar phase transitions occur in two-dimensional hard-core boson systems\cite{troyer}
 and in distorted frustrated
lattices.\cite{richter} 

\begin{figure}[b]
  \centering
  \includegraphics[width=.9\columnwidth,angle=0]{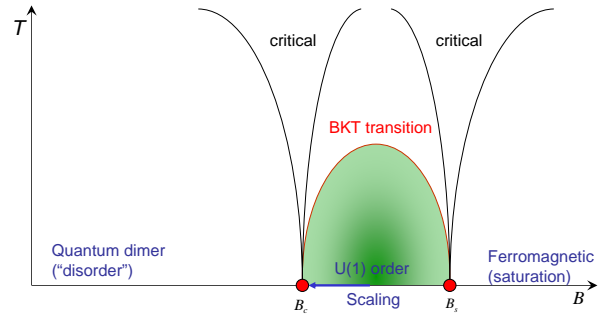} 
  \caption{(Color online) Schematic phase diagram.} 
  \label{scheme}
\end{figure}
The schematic behavior is illustrated in Fig.~\ref{scheme}. In this paper
we show that the crossover from BKT behavior to critical scaling is rather
well defined by the cooling rate and by characteristic maxima in the susceptibility.
However, this crossover occurs at distinctly higher temperatures than the 
BKT transition which can be determined 
by a careful analysis
of the spin-stiffness. There is no directly measurable signal for the BKT transition in
experiments,\cite{lang}  
but we find that magneto-caloric 
measurements are ideally suited to show the critical scaling and pinpoint the
exact location of the QPT.  
Close to the QPT the BKT transition retains the characteristic logarithmic 
behavior, albeit with strongly renormalized parameters.
We find, however, that the low temperature behavior above the QPT's 
does not fully follow theoretical expectations. 

\section{The model}\label{sec:model}
\begin{figure}[t]
  \centering
  \includegraphics[width=.6\columnwidth]{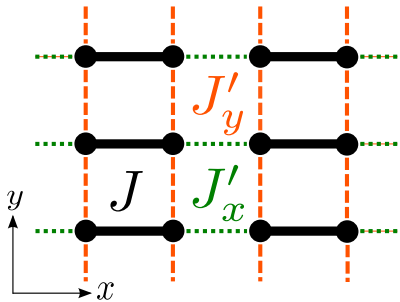}
  \caption{(Color online) Coupled dimers on a square lattice, with a columnar arrangement of the dimers.}
  \label{fig:squarelattice}
\end{figure}

%\subsection{Heisenberg model}
We use a ``columnar''  arrangement of strongly coupled antiferromagnetic 
dimers ($J>0$) on a two dimensional square lattice as shown in Fig.~\ref{fig:squarelattice},
described by the Hamiltonian of localized spin-1/2 operators $\hat{\vec{S}}_{x,y}$
\begin{align}
\begin{split}
\hat{\mathcal{H}} =& \sum_{y=1}^{N_y} \left[\, \sum_{x = \rm odd}^{N_x} J \hat{\vec{S}}_{x, y} \hat{\vec{S}}_{x+1, y}  + J'_x \hat{\vec{S}}_{x+1, y} \hat{\vec{S}}_{x+2, y}  \right. \\
&+\left.  J'_y \sum_{x=1}^{N_x} \hat{\vec{S}}_{x, y} \cdot \hat{\vec{S}}_{x, y+1} \right] - B \sum_{i=1}^{N} \hat{S}_{i}^{z},
\end{split} \label{eqn:HXXZ}
\end{align}
where the inter-dimer couplings $J'_x$ and $J'_y$ can be ferromagnetic or antiferromagnetic, 
but are assumed to be small $|J'| \ll J$.
%with
%\begin{itemize}
%\item $J'_x = \eta_x J$, $J'_y = \eta_y J$
%\item $\Delta_x = \Delta_y = 1$
%\item $B = g \mu_\text{B} B^{z}$
%\end{itemize}
%In this arrangement of the dimers it looks as if they are piled in columns, and because of this we call it a columnar dimer arrangement from now on.

\subsection{Interacting boson models}
Assuming that the intra-dimer exchange interaction $J$ dominates over 
inter-dimer couplings $J'_x$ and $J'_y$, it is natural to 
represent the system in the singlet and triplet basis at each dimer site
\begin{align}
\begin{split}
&\ket{t_-}_i = \kett{\downarrow \downarrow}_i\\
&\ket{t_0}_i = \frac{\kett{\uparrow \downarrow}_i + \kett{\downarrow \uparrow}_i}{\sqrt{2}}\\
&\ket{t_+}_i = \kett{\uparrow \uparrow}_i\\
&\ket{s}_i = \frac{\kett{\uparrow \downarrow}_i - \kett{\downarrow \uparrow}_i}{\sqrt{2}}.
\end{split}
\label{states}
\end{align}
At strong fields $B\approx J$ the last two states become nearly degenerate, 
while the other two higher energy states
will be neglected for now.  It is therefore justified to work in a  restricted Hilbert space 
with only two states at each dimer site, which are represented by hard-core
bosons on the vacuum $|0\rangle = \prod_i \ket{s}_i$ and
$b_j^\dagger |0\rangle = \ket{t_+}_j \prod_{i\neq j} \ket{s}_i$.
In this Hilbert space the 
effective Hamiltonian describes strongly interacting bosons on a rectangular lattice
\begin{eqnarray}
H_{\rm eff} & = & \sum_{\langle i,j \rangle} \left[-|t_{ij}| \left(b_i^\dagger b_j^{\phantom{\dagger} }
+ b_j^\dagger b_i^{\phantom{\dagger}}\right) + {t_{ij}} n_i n_j \right] \label{Heff1}\\
& & - \mu \sum_i n_i +
  U \sum_i n_i (n_i-1), \label{Heff}
\end{eqnarray}
where the limit $U\to \infty$ is implied to satisfy the hardcore constraint.
The effective chemical potential and the hopping in $x-$ and $y-$directions are given by 
\begin{eqnarray}
\mu  =   B-J, \  \ \
t_x  =  J'_x/4, \    \ \
t_y  =  J'_y/2 .
\end{eqnarray}
Note, that the hopping $|t_{ij}|$ in Eq.~(\ref{Heff1}) has been chosen to be positive, which can always be achieved by a local gauge transformation
$b_i \to (-1)^i b_i$. The nearest neighbor interaction in Eq.~(\ref{Heff1})
is repulsive (attractive)
for $J'>0$ ($J'<0$). 
By Fourier transforming the first term in the Hamiltonian the kinetic energy becomes
\begin{equation}
H_{\rm kin} =  \sum_{\vec{k}} (-2 |t_x| \cos k_x -2 |t_y| \cos{k_y})b_{\vec{k}}^\dagger b_{\vec{k}}^{\phantom{\dagger}}
\end{equation} 
The position of the upper and lower band edges allows 
a straight-forward estimate of the critical fields $B_c$ and $B_s$.
The lower critical field is determined by the chemical potential at which a single
boson  acquires positive energy $-2 |t_x|- 2 |t_y| =\mu$, which gives
\begin{equation} B_c \approx J-|J'_x|/2-|J'_y|. \label{bc} \end{equation}
This estimate is only correct to first order in $J'$, however, since the bosonic ground state
(vacuum) is not an exact eigenstate of the full Hamiltonian in Eq.~(\ref{eqn:HXXZ}).  
Higher order corrections from the neglected triplet states $\ket{t_-}$ and $\ket{t_0}$
  in Eq.~(\ref{states})
will be determined from numerical simulations as described below.

The upper critical field is determined from the energy gain of removing a particle from
the fully occupied band including the 
nearest neighbor interaction energy
\begin{equation} B_s = J+|J'_x|/2+|J'_y|+ J'_x/2+J'_y, \label{bs} \end{equation}
which is exact and corresponds to the saturation field of the original model (\ref{eqn:HXXZ}).
For intermediate fields $B_c< B < B_s$ the physics is governed by the behavior of 
two-dimensional interacting bosons (BKT phase) as explained below.

\subsection{The effective continuum model}\label{subsec:SCE}
We now focus on the lower QPT at $B_c$ which corresponds to 
the well studied case of a dilute interacting Bose gas.\cite{SachdevQPT}
At low filling the critical behavior of the lattice model in Eqs.~(\ref{Heff1})-(\ref{Heff})
is believed to be in the 
XY-universality class independent of the microscopic details.\cite{}
In the continuum limit the nearest neighbor interaction can be 
neglected and the hard-core constraint can be replaced by a strong $\phi^4$ interaction
of a complex bosonic field $\phi(\vec{r}, \tau)$ in a $(D+1)$-dimensional Euclidean action
\begin{align}
S =& \int\limits^{\beta}_{0}\text{d} \tau \int \text{d}^{D} r \left[ \overline{\phi}\left(\partial_{\tau} - \frac{\vec{\nabla}^2}{2m} - \tilde{\mu}\right)\phi + \frac{u_0}{2} \vert \phi \vert^4 \right]\text{,} \label{eqn:action}
\end{align}
where $D=2$ in our case.  
The parameters can be obtained by approximating the sums in Eqs.~(\ref{Heff1})-(\ref{Heff}) by 
integrals with a lattice spacing $a\approx 1$ and then rescaling $x'= x(t_y/t_x)^{1/4} , \ y'= y (t_x/t_y)^{1/4} $ 
\begin{eqnarray}
m = \frac{1}{2 a^2 \sqrt{|t_x t_y|}}; \ \ \ \tilde{\mu} = B-B_c;    \ \ \ u_0 = U a^2.
\label{rescaling}
\end{eqnarray}
In what follows we set the lattice spacing to unity $a=1$.

The action in Eq.~\eqref{eqn:action} describes an interacting dilute Bose gas 
with mass $m$ and chemical potential $\tilde\mu$.  For $\tilde\mu> 0$ or $B>B_c$ a finite
density of bosons appears even at zero temperature $T=0$, which signals the
QPT to the BKT phase.  Analogously, the same model also applies at the upper critical 
field $B_s$, where it describes bosonic singlet excitations on the saturated state with
$\tilde \mu = B_s - B$.

The upper critical dimensions is $D=2$ for this model so that 
logarithmic corrections appear in this case, 
which are described in terms of an ultraviolet cutoff 
$\Lambda_0$ (of the 
order of the reciprocal rescaled lattice spacing).   This situation ($D=2$) has been analyzed 
extensively in the literature\cite{PopovBook,PhysRevB.37.4936,PhysRevB.50.258,PhysRevB.73.085116,PhysRevB.59.14054,PhysRevA.66.043608,PhysRevB.69.144504,PhysRevLett.87.270402}
for various quantities which we summarize below.
Other dimensions are discussed in
the textbook of {Sachdev}.\cite{SachdevQPT}

The density of bosons 
corresponds to the magnetization per site
$\langle \overline{\phi}\phi\rangle = 2 M(B)/N$ in the spin dimer system as a function of field $\tilde{\mu} = B-B_c$, which is given at $T=0$ by\cite{PhysRevB.50.258}  
\begin{equation}
\frac{M}{N} =  \frac{m\tilde\mu\, \Theta(\tilde\mu)}{8\pi} \ln\left[ \frac{\Lambda^2_0}{2m\tilde\mu} \right] \text{.} \label{MB}
\end{equation}
The susceptibility is therefore
\begin{align}
\chi  =&\: \frac{m}{8\pi}\left(\ln\left[ \frac{\Lambda^{2}_{0}}{2m\tilde\mu}\right] - 1 \right)\label{eqn:suscyboson2d}\,,
\end{align}
which is logarithmically divergent as the critical point is approached from above 
inside the BKT phase $B\to B_c$.
For $T> 0$ and $B=B_c$ it has been predicted that the density increases with temperature
including a characteristic logarithmic correction\cite{PhysRevB.50.258}
\begin{equation}
M(T) = \frac{m  T}{4 \pi} \ln^{-4}\left[\frac{\Lambda^2_0}{2 m  T}\right].\label{MT}
\end{equation}
The scaling as a function of $T$ can be used 
in order to identify the exact value of the critical field $B_c$ as outlined 
below.

Finally, the BKT transition temperature has been predicted as a function of field\cite{PhysRevB.37.4936}
\begin{align}
T_{\rm BKT} = \frac{\tilde \mu}{4} \frac{\ln\left[ \frac{\Lambda_0^2}{2 m\tilde\mu} \right] }{\ln\left[ \ln\left[ \frac{\Lambda_0^2}{2 m\tilde\mu} \right]  \right]}\label{eqn:TBKT}\,.
\end{align}
However, for this formula to be valid the double logarithm has to become very large, 
which does not correspond to physically relevant regions.\cite{PhysRevB.37.4936,PhysRevB.73.085116}
In fact, it remains to be seen if the single logarithms in Eqs.~(\ref{MB})-(\ref{MT}) are large enough so that the leading behavior can be observed in our numerical simulations below and 
in future experiments.

\section{Determining the critical fields}\label{sec:CriticalFields}

In order to analyze the quantum phase transitions, the exact locations 
of the critical fields have to be determined
first.  
As mentioned above, the upper critical field $B_s$ is exactly the saturation field
in Eq.~(\ref{bs}), but the lower field in Eq.~(\ref{bc}) 
will in general have higher order corrections of the form
\begin{equation} B_c \approx J-|J'_x|/2-|J'_y| + a_x J'^2_x + a_y J'^2_y + a_{xy} J'_x J'_y. 
\label{bc2} \end{equation}
The higher order corrections are due to 
virtual excitations to the neglected triplet states $\ket{t_-}$ and $\ket{t_0}$
  in Eq.~(\ref{states}).
The exact values for $a_x = - 0.375$ and $a_y = 0.5 J_y$ are known from 
higher order strong coupling expansions for the 
dimerized chain\cite{PhysRevB.59.11384, PhysRevB.7.3166} $(J'_y=0)$ 
and for the ladder system\cite{Rotzki} $(J'_x=0)$, respectively.

\begin{figure}[t]
  \centering
  {
  \includegraphics[width=0.8\columnwidth]{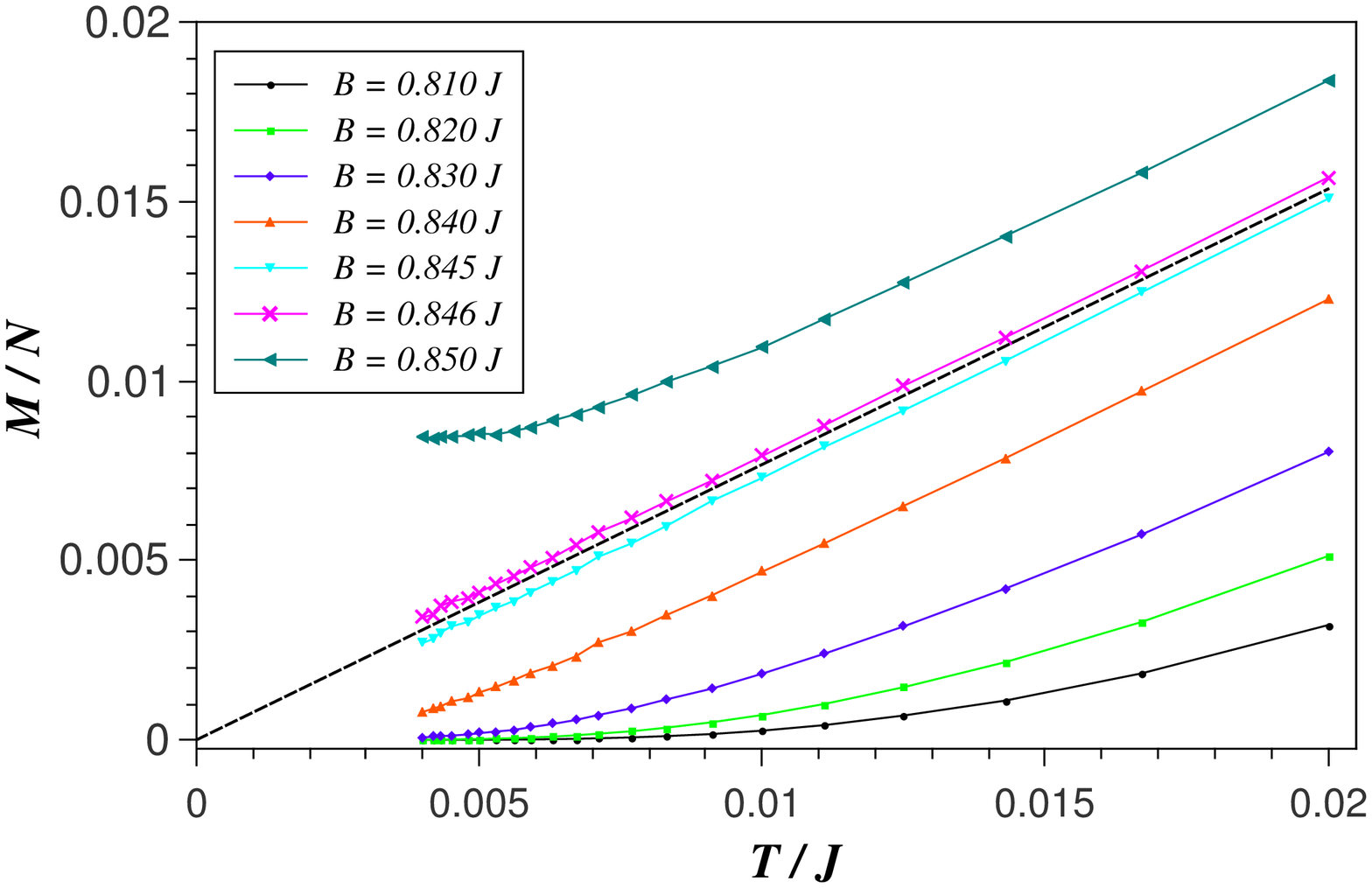}\vspace{0.5cm}\\
  \includegraphics[width=0.8\columnwidth ]{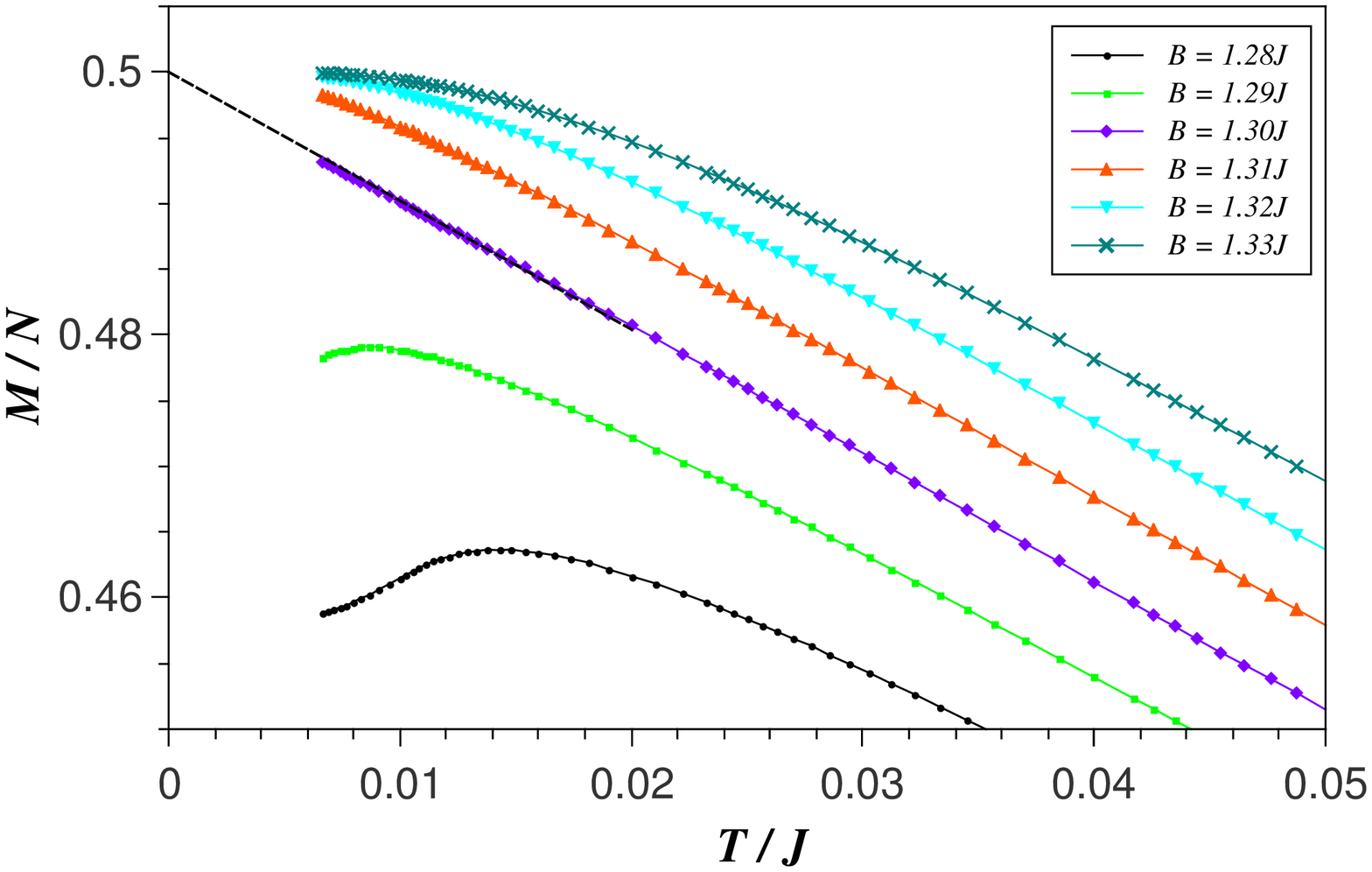}
  }
  \caption[Magnetization as a function of temperature for different magnetic fields]{(Color online) Magnetization as a function of temperature for different magnetic fields for 
for $J'_x=J'_y = 0.1$ and $N = 676$ near $B_c$ (top) and  $B_s$ (bottom).}
  \label{fig:MagT26}
\end{figure}

In order to determine the exact location of the QPT for   general inter-dimer  couplings, 
numerical simulations
at $T=0$ in the thermodynamic limit would be required.
This is obviously impossible, 
but large systems sizes at small finite temperatures are feasible
with Quantum Monte Carlo (QMC) simulations.  In order to examine the model in 
Eq.~(\ref{eqn:HXXZ}) numerically, we therefore have implemented the Stochastic
Series Expansion algorithm\cite{Sandvik2002} with directed loop updates and using
the so-called
Mersenne Twister random number generator.\cite{Mersenne1998}

At finite temperatures the discontinuity in Eq.~(\ref{MB}) cannot be observed directly, 
but the magnetization
as a function of temperature 
becomes exponentially small for $B<B_c$ while it approaches a finite value for $B>B_c$.
The critical field $B_c$ is then exactly defined as the point where 
   critical scaling is obeyed, which can be determined
rather accurately.  
This behavior is illustrated in Fig.~\ref{fig:MagT26}.

 \begin{table}[b]
 \begin{center}
 \begin{tabularx}{0.9\columnwidth}{c|c|c|c|c}
 \hline\hline
   case & $t_x$ $[J]$ & $t_y$ $[J]$ & $B_c\  [J]$ & from Eqs.\\
& & &$\pm 0.0005$ & (\ref{bc2}), (\ref{coeff})  \\
  \hline
  $J'_x = J'_y=0.1 J$ & $0.025$ & $0.05$ & $0.8460$ &  0.84625 \\
    %\hline
  $J'_x = 2J'_y=0.15 J$ & $0.0375$ & $0.0375$ & $0.8391$ & 0.83875   \\
    %\hline
  $2J'_x = J'_y= 0.12J$ & $0.015$ & $0.06$ & $0.8523$  & 0.85225 \\
  \hline\hline
 \end{tabularx}
 \caption{Critical field $B_c$ for three different choices of exchange couplings, which 
obey the condition 
$J'_x + 2J'_y = 0.3J$, i.e.~$B_c \approx 0.85 J$ to lowest order.}\label{table:hoppings}
 \end{center}
 \end{table}

Note, however, that the observed scaling in 
Fig.~\ref{fig:MagT26} at the exactly known upper critical field $B_s$ appears to be
perfectly linear (relative to the saturated state).   This means that the logarithmic 
correction in Eq.~(\ref{MT}) must be very small, which puts a lower limit on
the cutoff $\Lambda_0 \agt 10^7$.  To determine the lower critical field $B_c$, we therefore
use linear scaling as well.
Extrapolating the data to the thermodynamic limit and then determining the
critical fields $B_c$ by the best linear fit 
gives the results for three different choices of inter-dimer couplings
shown in Table~\ref{table:hoppings}.
Ignoring higher orders, the values for the coefficients in Eq.~(\ref{bc2})
are then consistent with the following estimates
\begin{equation}
a_x = -0.375, \ \ a_y =  0.5, \ \ a_{xy} \approx -0.5 \pm 0.03  \label{coeff}
\end{equation}

\begin{figure}[t]
  \centering
  \includegraphics[width=0.9\columnwidth]{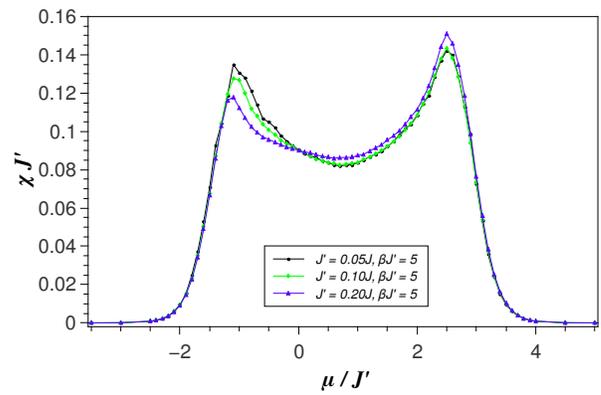}
  \caption{(Color online) The susceptibility $\chi J'$ as a function of $\mu/J'= (B-J)/J'$
at inverse temperature $\beta J'= 5$ for three inter-dimer coupling strengths 
$J'_x = J'_y = J' = 0.05 J$,
$0.1 J$ and $0.2J$ }
  \label{fig:univers1}
\end{figure}

Before continuing our analysis we would also like to consider how the neglected  
higher energy triplet excitations $\ket{t_-}$ and $\ket{t_0}$
  in Eq.~(\ref{states}) affect physical observables like the magnetization.
We note that the effective Hamiltonian
(\ref{Heff1})-(\ref{Heff}) is invariant under changing the inter-dimer coupling strengths
$J'_x$ and $J'_y$ as long as all energies and the field $\mu$ are rescaled
accordingly.
We therefore consider three different realization of the coupling strength
$J'_x = J'_y = J' = 0.05 J$,
$0.1 J$, $0.2J$ and plot the 
susceptibility $\chi J'$ 
as a function of rescaled field $\mu /J'= (B-J)/J'$
at a given rescaled temperature $\beta J'= 5$  
in Fig.~\ref{fig:univers1}.
We observe a finite susceptibility in the BKT phase with two characteristic maxima near
the QPT.
While the three
curves agree reasonably well, systematic deviations can be seen for larger $J'$, which 
can only come due to 
corrections from the higher energy triplet excitations. 
In what follows we choose a coupling strengths of $J'_x = J'_y = J' =  0.1 J$, 
which is a compromise between minimizing higher order corrections and efficient numerical
simulations.  It is believed that the higher order triplet excitations do not change
the form of the critical scaling in Eqs.~(\ref{MB})-(\ref{eqn:TBKT}).

\section{Critical scaling at the QPT}\label{sec:ScalingB}
\begin{figure}[t]
  \centering
  {
  \includegraphics[width=0.9\columnwidth]{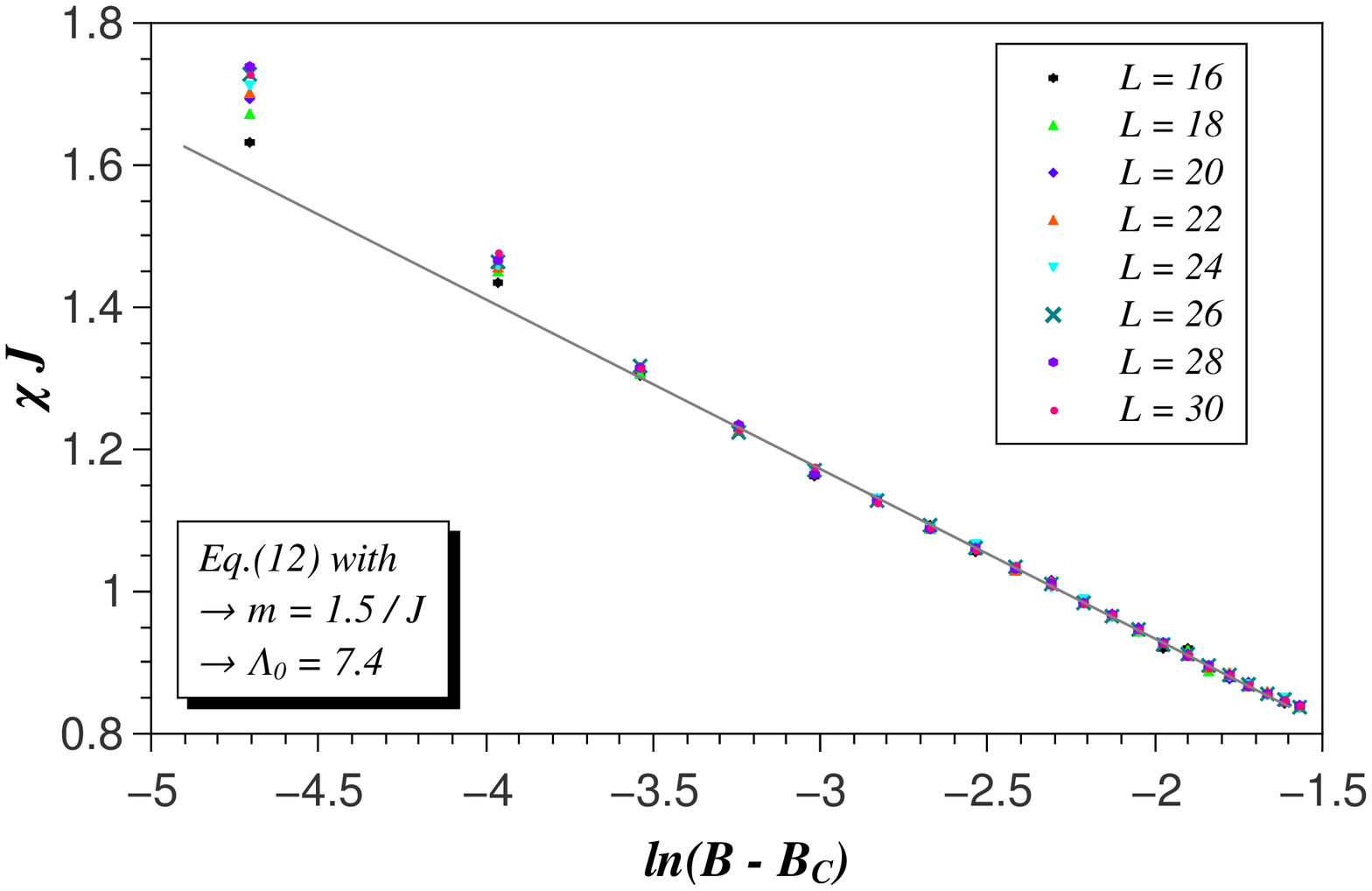}\\
  \vspace{0.15cm}
  \includegraphics[width=0.9\columnwidth]{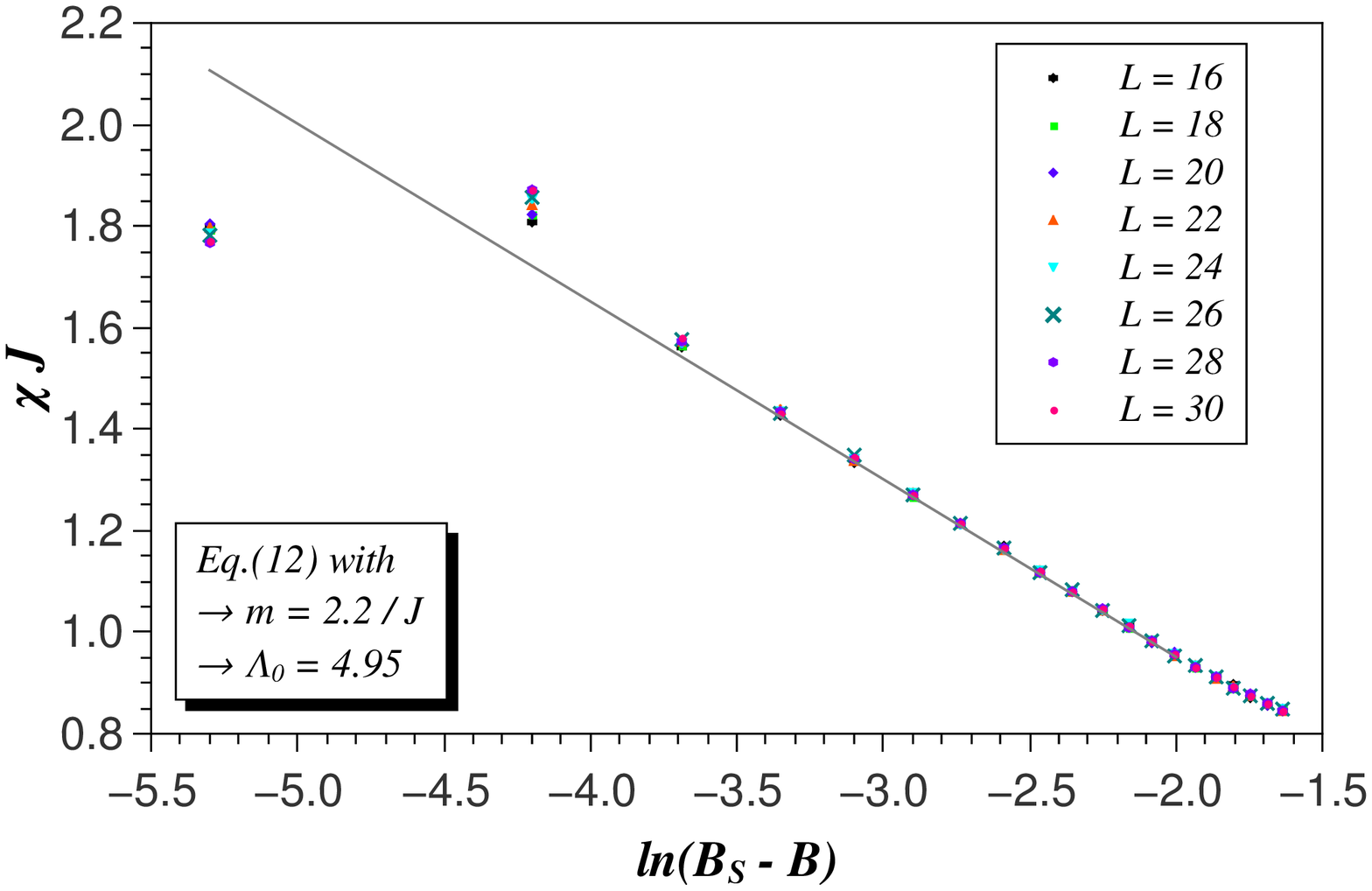}
  }
  \caption[Susceptibility for different system sizes $N$ and an inverse  temperature of $\beta J = 200$ for $J' = 0.1J$]{(Color online) Susceptibility for different system sizes $N$ and an inverse temperature of $\beta J = 200$ for $J'_x = J'_y = J' = 0.1J$ near $B_c$ (top) and $B_s$ (bottom).}
  \label{fig:Scaling1}
\end{figure}

\begin{figure}[t]
  \centering
  {
  \includegraphics[width=0.9\columnwidth]{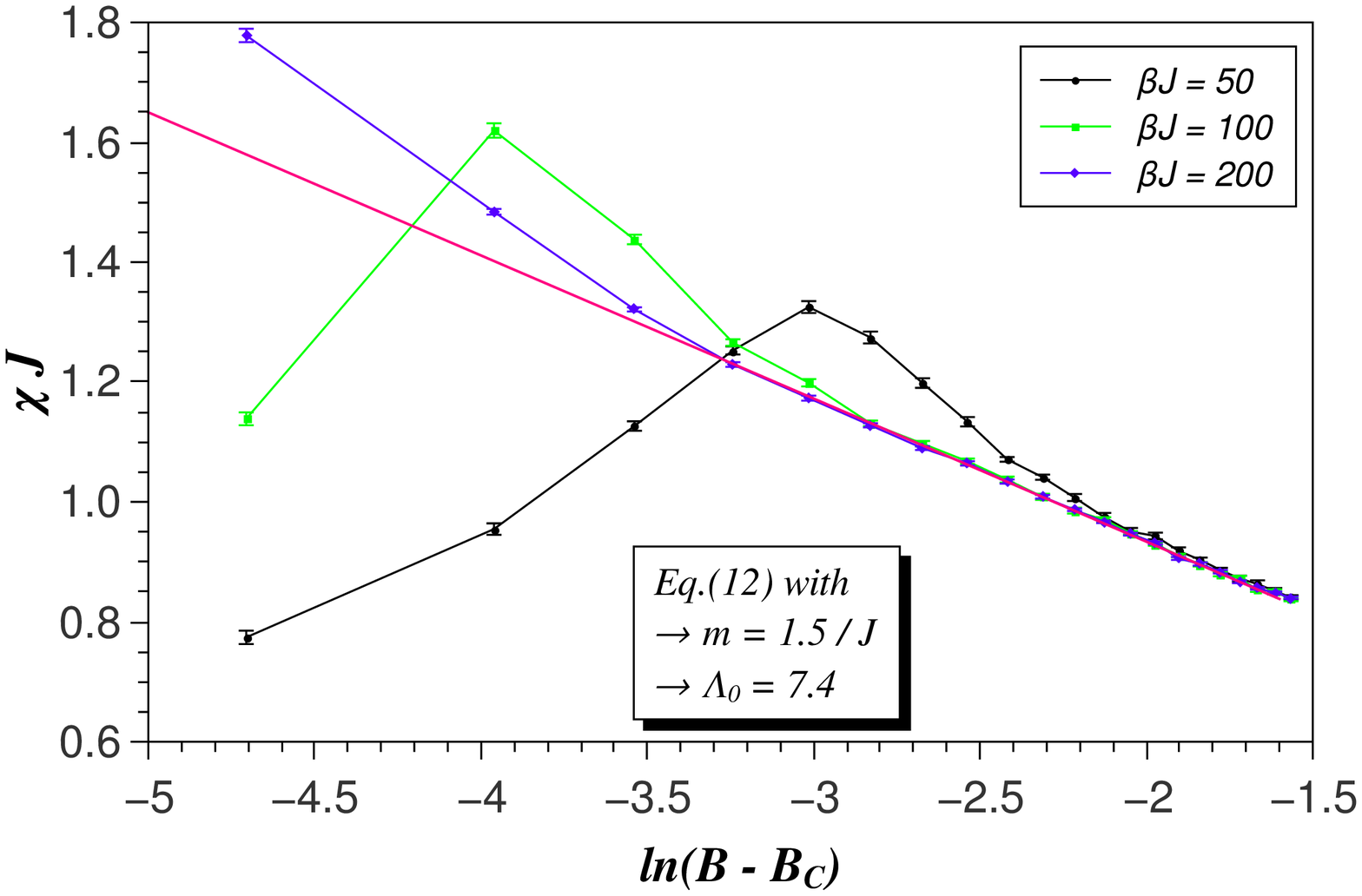}\\
  \vspace{0.15cm}
  \includegraphics[width=0.9\columnwidth]{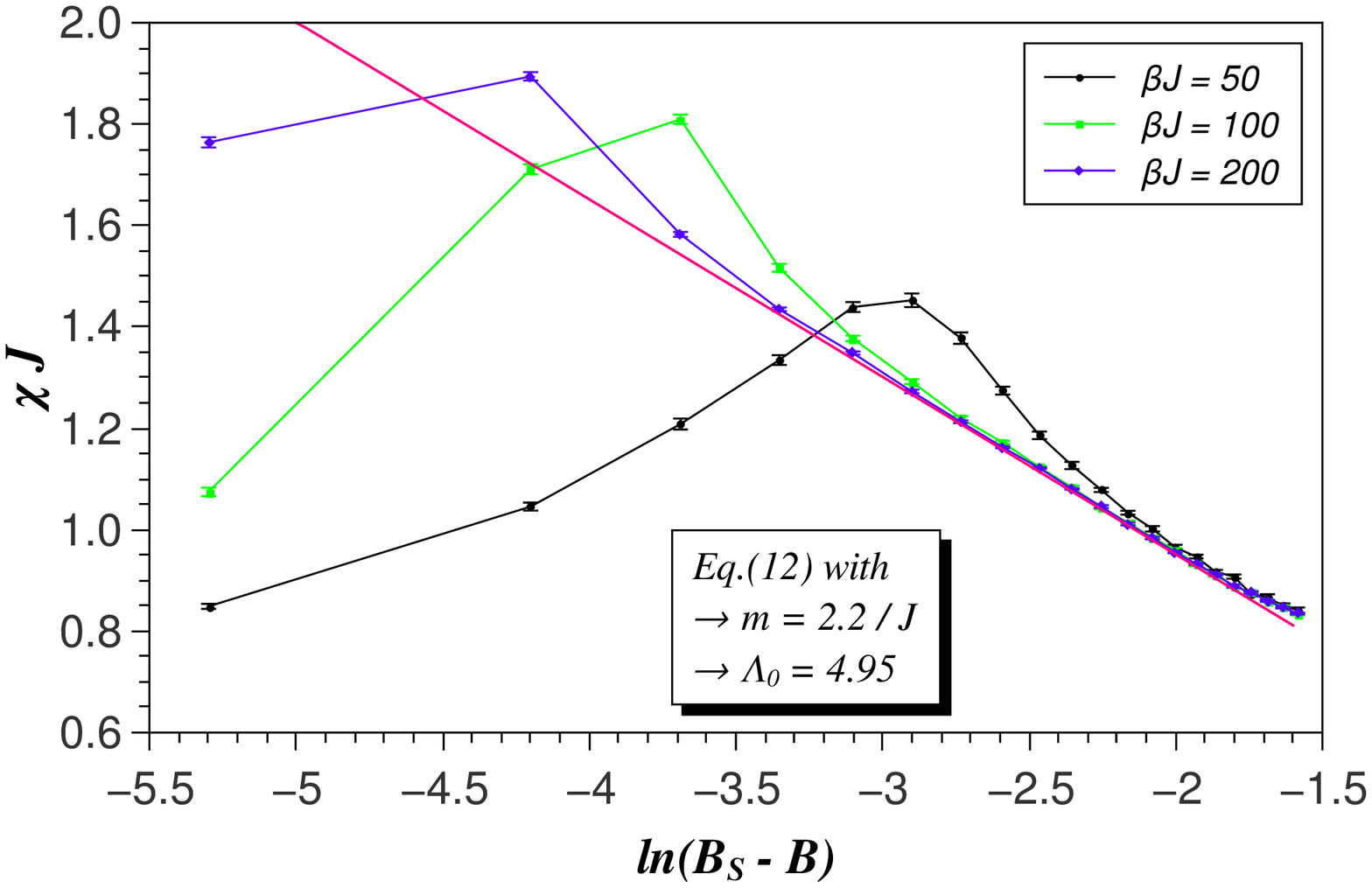}
  }
  \caption[Susceptibility in the thermodynamic limit for different inverse temperatures $\beta J$ and $J'_x = J'_y = J' = 0.1J$]{(Color online) Susceptibility in the thermodynamic limit for different inverse temperatures $\beta J$ and $J'_x = J'_y = J' = 0.1J$ near $B_c$ (top) and $B_s$ (bottom).}
  \label{fig:Scaling2}
\end{figure}

We now turn to analyzing the scaling behavior of the susceptibility $\chi$ 
as a function
of field $B$ in Eq.~(\ref{eqn:suscyboson2d}).  Finite temperatures $T$ and 
system sizes $N=L\times L$
play the role of an infrared cutoff $D_0 \sim {\rm max}(T,J'/L)$ which 
will give deviations from the predicted
$T=0$ scaling in Eqs.~(\ref{MB}) and (\ref{eqn:suscyboson2d})
as the QPT is approached.  However, for fields $|B-B_{c/s}|  \agt D_0$
the scaling can still be tested.  
At each given temperature we first 
increase the system size until systematic convergence of the magnetization is obtained 
as shown in Fig~\ref{fig:Scaling1}.   
The resulting susceptibility in the thermodynamic limit
near the QPT's is shown in Fig.~\ref{fig:Scaling2}
as a function of 
the logarithm of $\tilde \mu = |B-B_{c/s}|$ 
for different 
temperatures.


The data confirms that the scaling approaches a logarithmic behavior for $T\to 0$ 
consistent with the form in Eq.~(\ref{eqn:suscyboson2d}). 
We notice that the finite temperature
susceptibility is actually rather small at the QPT $B=B_c$, but then
increases and {\it overshoots} the logarithmic divergence, before 
the logarithmic behavior is reached inside the BKT phase.
In this way the field-integral of the susceptibility (i.e.~the magnetization) 
remains largely temperature independent outside the critical region, 
since the smaller values at the QPT for finite $T$ are  compensated 
by a corresponding
overshooting of the maximum.
%Because of this the magnetization is largely temperature independent 
%deep inside the BKT phase. 
In turn this means that the 
characteristic maxima in Fig.~\ref{fig:univers1} 
of the susceptibility are only indirectly related to the QPT. The overshooting implies
that the large fluctuations in the magnetization arise from a different mechanism 
at finite temperatures.  One may expect that the maxima are therefore related to the
finite-temperature BKT transition, but this is {\it not} the case as we will see below.
Instead we find that the susceptibility maxima are found for temperatures
well above the BKT transition $T>T_{\rm BKT}$ at the corresponding fields. 
As we will see later the maxima
coincide with maxima in the entropy, so that  
these points correspond  to 
the crossover between quantum critical scaling to 
vortex physics.

Comparing with the expected form in Eq.~(\ref{eqn:suscyboson2d}) quantitatively, we
find rather small values of
the effective mass $m \approx 1.5/J $ at the lower QPT $B_c$
and $m\approx 2.2/J$ at $B_s$, which are strongly renormalized 
compared to the naive estimate  $m\approx 14/J$  according 
to Eq.~(\ref{rescaling}).   
The value of 
$\Lambda_0 \sim 5-7$ remains finite in Eq.~(\ref{eqn:suscyboson2d}).
The value of $m$  from the fits at the lower QPT 
is rather sensitive to the exact location of the critical field 
$B_c$. In general all microscopic details
such as the 
neglected next-nearest neighbor interaction in Eq.~(\ref{Heff1}) will 
influence the exact value of the effective parameters in Eq.~(\ref{rescaling}).

\section{The Berezinsky-Kosterlitz-Thouless phase transition}\label{sec:BKT}

The intermediate region between the two QPT's is dominated by the 
presence of interacting triplon excitations which form a condensate at $T=0$ with
long range phase coherence.   We now consider the finite temperature behavior in this
intermediate phase.  While the QPT's are driven by 
quantum fluctuations, the transition due to thermal fluctuation corresponds to 
classical behavior and is therefore not directly related to the scaling discussed above. 

\begin{figure}[t]
  \centering
  {
  \includegraphics[width=.95\columnwidth]{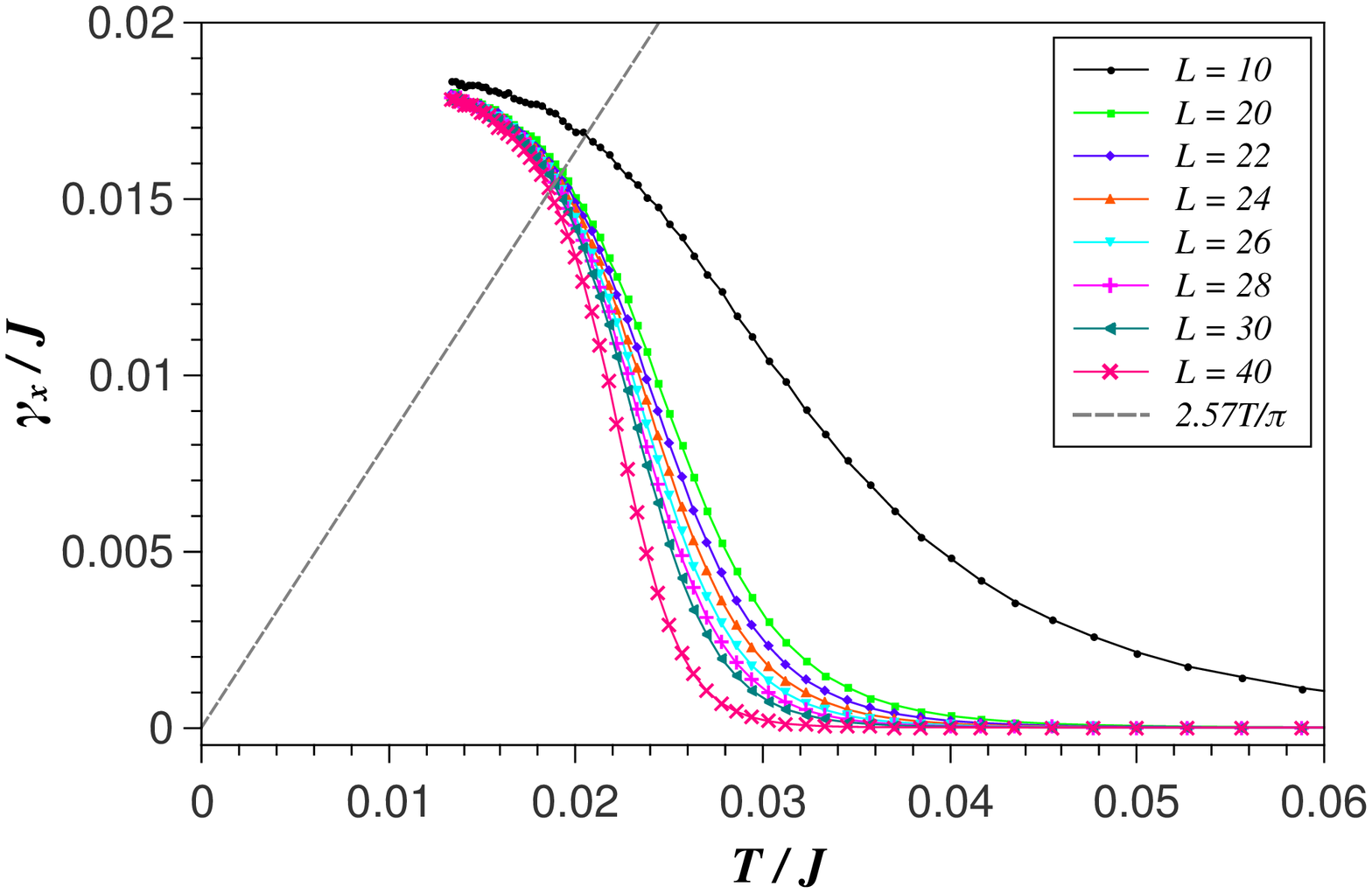}\\
  \ \ \ \includegraphics[width=0.85\columnwidth]{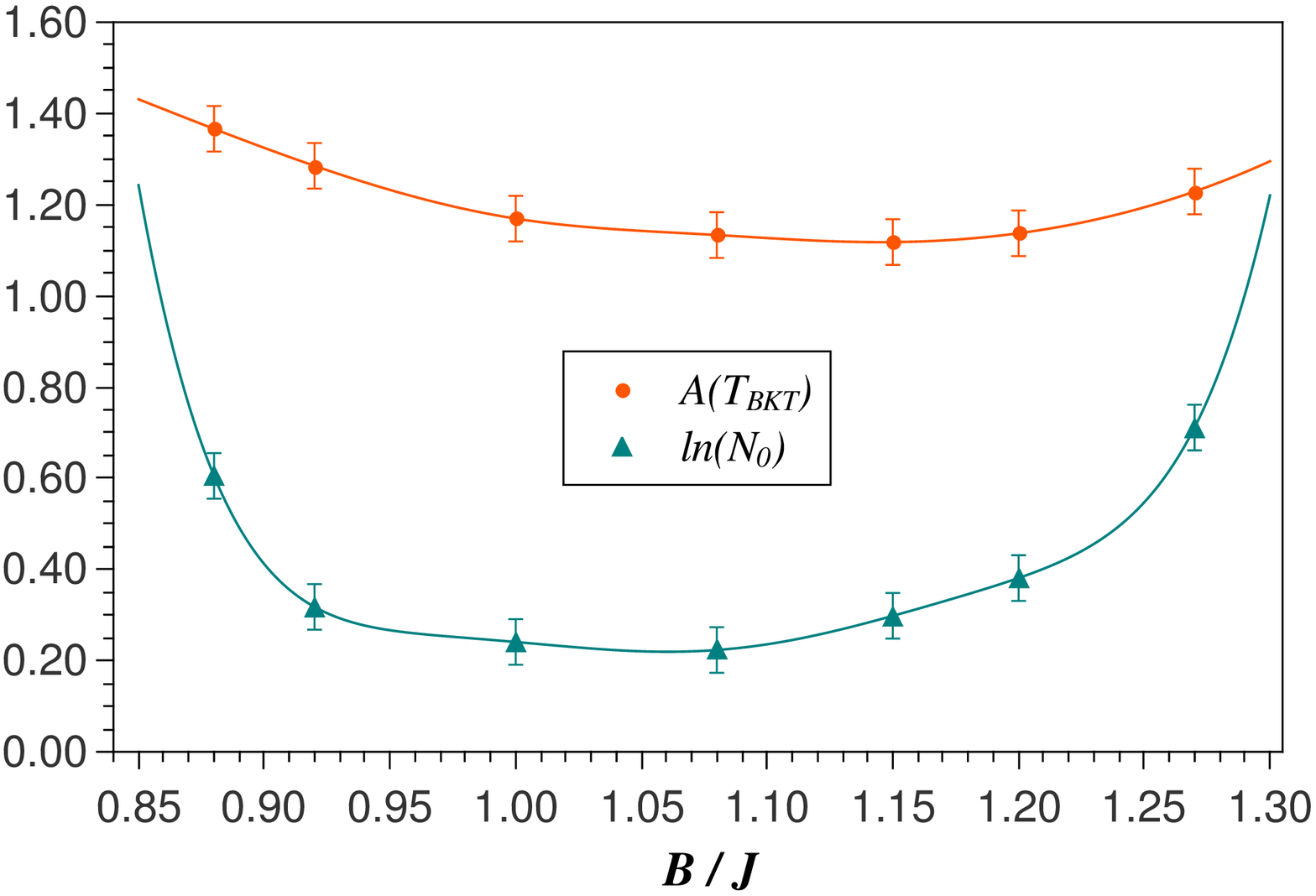} 
  }
  \caption{(Color online) Top: Helicity modulus for different system sizes as a function of temperature at
$B=0.92J$.
Bottom: $A(T_{\rm BKT})$ and $\ln(N_0)$  as a function of field. }
  \label{fig:helicity}
\end{figure}

The effective hard-core boson 
model in Eqs.~(\ref{Heff1})-(\ref{Heff}) is exactly equivalent to the
XXZ-spin model with $J_z=J_{xy}/2$, which is known to be in the XY-universality class.
At finite temperatures this system undergoes a BKT transition, which can be 
described in terms of classical two-dimensional spins as first explained in the works
of {Berezinsky}{\cite{Bere1,Bere2}} and {Kosterlitz} 
and {Thouless}.{\cite{KT1,KT2}} 
At low temperatures $T<T_{\rm BKT}$ a quasi long-range ordered phase with 
power-law decay of correlations exists. Above the phase transition 
temperature $T_{\rm BKT}$ the unbinding of vortices is energetically 
allowed leading to a disordered phase with exponential decaying correlations.
{Kosterlitz} used the spin stiffness\cite{KT2}
\begin{align}
\rho_S = \left. \frac{1}{N} \left( \frac{\partial^2 F}{\partial \phi^2} \right)\right\vert_{\phi=0} \label{eqn:sfdensity}
\end{align}
to identify a phase transition, where $F$ is the free energy of the system and $\phi$ 
is the angle between spins at opposite edges of the system.
In order to determine the spin stiffness in 
QMC simulations it is convenient to calculate the winding number fluctuations
in each direction,\cite{PhysRevB.36.8343,PhysRevB.56.11678}
which can be used to define a so-called helicity modulus\cite{helicity}
\begin{align}
\gamma =&\, \frac{T}{D} \left\langle \vec{\upomega}^2 \right\rangle =\, \frac{T}{2} \left( \left\langle \upomega^{2}_{x} \right\rangle + \left\langle \upomega^{2}_{y} \right\rangle \right)  \label{eqn:definehelicity} \text{.}
\end{align}
The phase transition temperature $T_{\rm BKT}$
is then determined by the value where the helicity
modulus obeys\cite{KT2}
\begin{align}
\gamma(T_{\rm BKT}) =&\, \frac{\hbar^2}{m^2} \rho_S(T_{\rm BKT}) =\, \frac{2}{\pi}T_{\rm BKT} \label{eqn:helidrop}.
\end{align}
Generally the helicity modulus has a more complex relation to the spin-stiffness
for anisotropic systems in low dimensions $D$ as discussed by {Prokof'ev} and {Svistunov}.\cite{PhysRevB.61.11282}  For anisotropic systems it is therefore useful to define
a separate helicity modulus for each direction\cite{PhysRevB.69.014509}
\begin{align}
\begin{split}
\gamma_x =\,& T  \frac{L_x}{2t_x} \frac{t_y}{L_y} \left\langle \upomega^{2}_{x} \right\rangle \\
\gamma_y =\,& T \frac{L_y}{t_y} \frac{2t_x}{L_x} \left\langle \upomega^{2}_{y} \right\rangle,
\end{split}\label{eqn:helicitydiffdir2} 
\end{align}
where $L_x/2$ and $L_y$ are the edge lengths of the effective hard-core boson system in terms
of the size of the original spin system $N=L_x\times L_y$.  Instead of taking 
the average in Eq.~(\ref{eqn:definehelicity}), only the largest one contributes
$\gamma= {\rm max}({\gamma_x,\gamma_y}) =\gamma_x$,
while the smaller one shows a linear behavior with edge length 
$\gamma_y\propto L_y$.\cite{PhysRevB.69.014509,PhysRevB.61.11282}
\begin{figure}[t]
  \centering
  {
  \includegraphics[width=0.9\columnwidth]{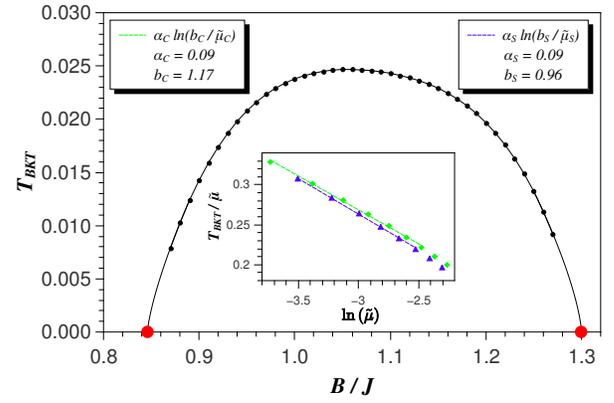}\\
  %\includegraphics[width=0.9\columnwidth]{TBKT_scaleBC_helicity_x.eps}\\
  %\vspace{0.15cm}
  %\includegraphics[width=0.9\columnwidth]{TBKT_scaleBS_helicity_x.eps}
  }
  \caption[Scaling of the critical BKT temperature for $N = \infty$ and $J' = 0.1J$]{(Color online) The 
BKT temperature as a function of field.   Inset: Logarithmic behavior according to Eq.~(\ref{BKT2}) near the critical fields.}
  \label{fig:TBKT}
\end{figure}

The energy of the vortices also depend logarithmically on the system size $N$, so that 
the condition in Eq.~(\ref{eqn:helidrop}) acquires a corresponding 
correction\cite{PhysRevB.37.5986} 
\begin{align}
\begin{split}
\frac{\pi \gamma_x(N,\, N_0)}{2 T} =& A(T) \left( 1 + \frac{1}{2}\cdot\frac{1}{\ln(N/N_0)} \right),
 \label{eqn:BKTfit2}
\end{split}
\end{align}
where $N_0$ is a fitting parameter and 
$A(T)$ should take on the universal value of unity at the transition, but 
can also be used as a fitting parameter.\cite{JPSJ.67.2768,PhysRevB.67.104414}
Following the procedure in Ref.~[\onlinecite{JPSJ.67.2768}] the logarithmic corrections in
Eq.~(\ref{eqn:BKTfit2}) become only accurate at the phase transition, which 
can in fact be used to determine $T_{\rm BKT}$ and $A(T_{\rm BKT})$.  
In Fig.~\ref{fig:helicity} the helicity modulus is plotted at a given field $B=0.92J$ for
different system sizes.  
The BKT transition for each field is 
determined by the best fit of Eq.~(\ref{eqn:BKTfit2}), i.e.~when $\pi \gamma_x(N,\, N_0)/2 T$ extrapolates to a limiting value linearly as
a function of $\ln^{-1}(N/N_0)$.  For a classical isotropic 
spin model a value of $A(T_{\rm BKT})=1$ 
can be confirmed,\cite{JPSJ.67.2768,bkt-sandvik} 
but for the spin dimer model we find a field-dependent
value for $A(T_{\rm BKT})$ which is 
slightly larger than unity as given in Table \ref{table:gammainflnl0} and shown in 
Fig.~\ref{fig:helicity}.  The fitting parameter $N_0$
also becomes field dependent.  The resulting transition temperature is shown in 
Fig.~\ref{fig:TBKT}, which shows a sharp drop near the QPT.
As shown in the inset the behavior is consistent with a logarithmic 
behavior 
\begin{equation}
\frac{T_{\rm BKT}}{ \tilde \mu} \approx \alpha \ln (b/\tilde\mu), \label{BKT2}
\end{equation}
but the double logarithmic correction in the asymptotic scaling at extremely small 
densities in Eq.~(\ref{eqn:TBKT}) cannot be 
confirmed numerically.\cite{troyer}  

 \begin{table}[b]
 \begin{center}
 %\begin{tabularx}{0.98\columnwidth}{c|c|c|c}
 \begin{tabular}{c|c|c|c}
    \hline\hline
  $B [J]\:\:$ & $A(T_{\rm BKT}) \pm 0.03\:\:$ & $\ln\left( N_0 \right) \pm 0.05\:\:$ & $T_{BKT}/J \pm 0.0005$\\
    \hline
  $0.880$ & $1.37$ & $0.61$ & $0.0103$ \\
  $0.920$ & $1.29$ & $0.32$  & $0.0174$ \\
  $1.000$ & $1.17$ & $0.24$ & $0.0239$ \\
  $1.080$ & $1.14$ & $0.22$ & $0.0245$ \\
  $1.150$ & $1.12$ & $0.30$ & $0.0229$ \\
  $1.200$ & $1.14$ & $0.38$ & $0.0197$ \\
  $1.270$ & $1.23$ & $0.71$ & $0.0092$ \\
     \hline\hline
 \end{tabular}
 \caption{Results of $A(T_{\rm BKT})$ and $\ln\left( N_0 \right)$ for different magnetic fields.}
\label{table:gammainflnl0}
 \end{center}
 \end{table}
The deviations from $A(T_{\rm BKT})=1$ can be traced to two different
sources:  In the middle of the BKT phase we find that a nearly isotropic effective system
with $L_x=2 L_y$ and $J'_x = 2 J'_y$ gives values of $A(T_{\rm BKT}) \approx 1.04$,
so that the detailed geometry appears to have some effect on the exact value of $A$.
A second source may be higher order corrections
in $\ln N/N_0$, which can be expected to become significant when the effective density of 
bosons per lattice site is small, which in turn leads to large distances between vortices. 
Therefore, the corrections must be largest
close to the QPT, consistent with our findings.  
Using a constant value of $A(T_{\rm BKT})=1$ in the fits changes the estimate for
$T_{\rm BKT}$ by up to 10-15\%.

\begin{figure}[t]
  \centering
  \includegraphics[width=\columnwidth]{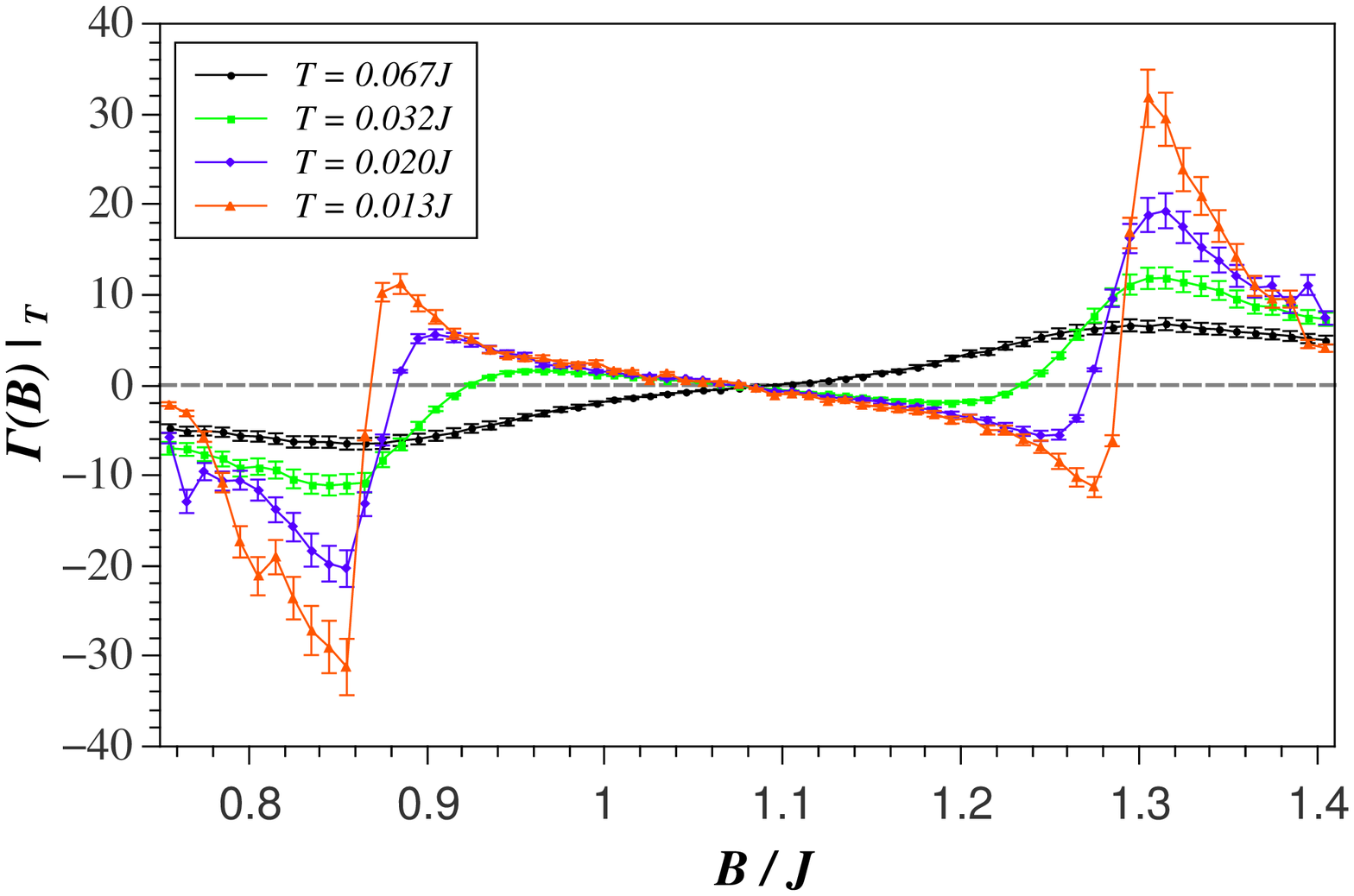} 
  \includegraphics[width=\columnwidth]{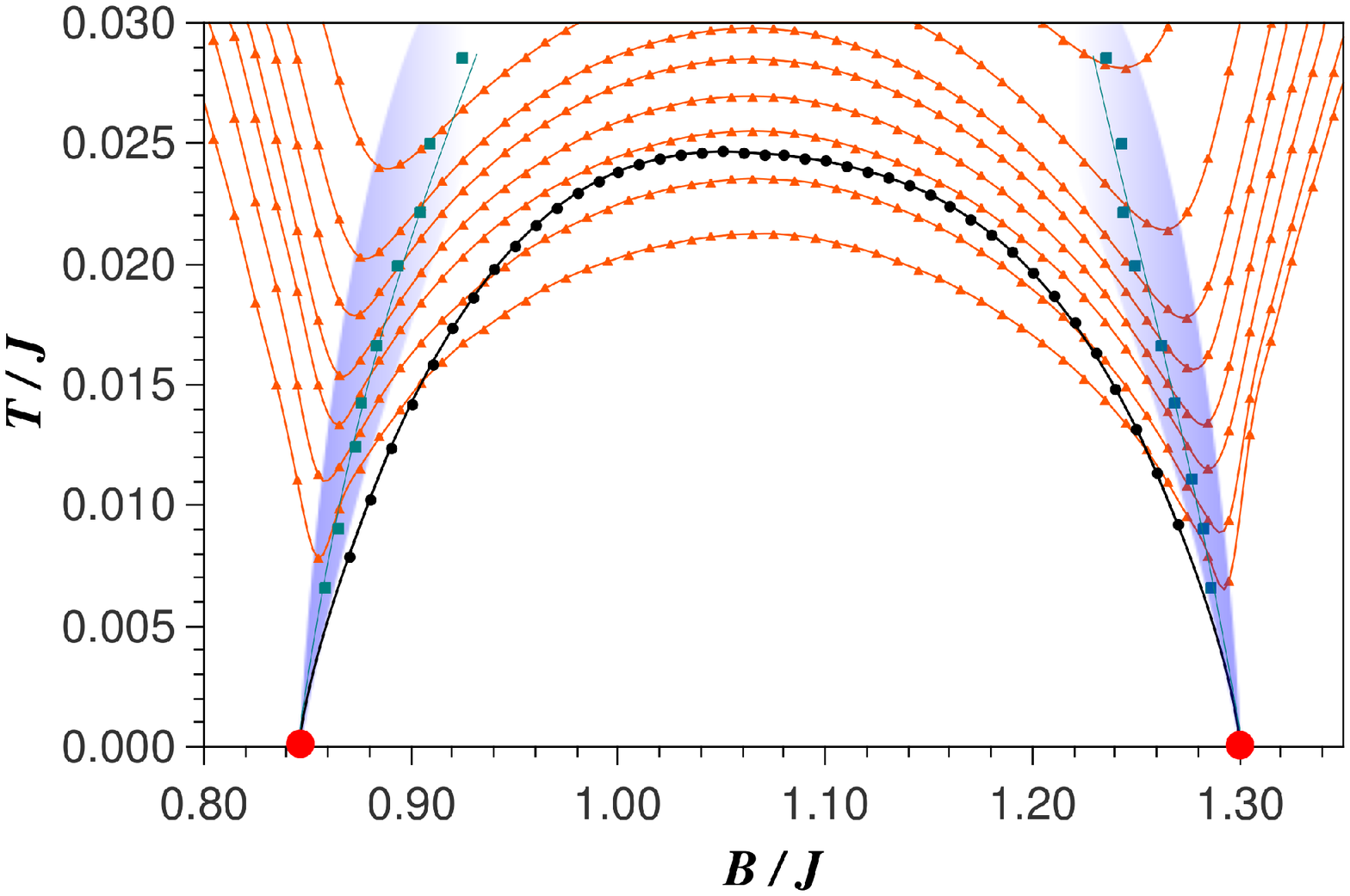}
  \caption[$T$-$B$ phase diagram for the dimerized columnar square lattice]{(Color online) 
Top: Cooling rate $\Gamma$ as a function of field for different temperatures.
Bottom: Temperatures as a function of field for different values of constant entropy
(red isentropes $dS=0$) near $B_c$.  The shaded region is dominated by
large entropy, corresponding to the
minima in the isentropes, which are relatively close to 
the maxima in the 
susceptibility (marked by green squares). 
The BKT transition $T_{\rm BKT}$  (connected dots)  
occurs at significantly lower temperatures.  }
 \label{fig:gamma}
\end{figure}
\section{Magneto-calorics and the $T$-$B$ phase diagram}
As we already discussed in Sec.~\ref{sec:CriticalFields},
the behavior of the magnetization $M(T)$ as a 
function of temperature plays an important role in determining the locations of the
QPT.  The interplay of magnetization with temperature is often termed
magneto-calorics, which 
has been a fruitful field
ever since the discovery of adiabatic demagnetization by
{Warburg} in 1881.\cite{Warburg1881}
The central quantity of interest in this context is
the cooling rate
\begin{align}
\begin{split}
\Gamma(B,T) = \frac{1}{T} \left( \frac{\partial T}{\partial B} \right)_{\text{S}}   
\end{split}\label{Gamma} \,,
\end{align}
which describes the temperature change with the applied field under adiabatic conditions.
Using the cyclical rule and a Maxwell relation the cooling rate is also directly 
related to $M(T)$ and $S(B)$
\begin{align}
\begin{split}
\Gamma(B,T) 
= - \frac{1}{C} \left( \frac{\partial S}{\partial B} \right)_{\text{T}}
= - \frac{1}{C} \left( \frac{\partial M}{\partial T} \right)_{\text{B}}
\end{split}\label{eqn:MCE} \,.
\end{align}
where 
$C = T \left( \frac{\partial S}{\partial T} \right)_{\text{B}}$ is the heat capacity.
Therefore, the entropy is largest when $\Gamma =0$.

The cooling rate for different temperatures is plotted in Fig.~\ref{fig:gamma} (top), 
which shows
sharp features near the QPT.
In Ref.~[\onlinecite{PhysRevB.72.205129}] it was predicted that the
cooling rate diverges with a universal prefactor near the QPT, but we were not 
able to reach low enough temperatures to confirm this behavior.

Integrating the cooling rate in Eq.~(\ref{Gamma}) gives the temperature as a
function of field for a given entropy $S$.   The corresponding 
isentropes are  shown in Fig.~\ref{fig:gamma} (bottom).
The temperature reaches a minimum when the cooling rate is zero, which means that the
entropy as a function of field (horizontal path) is maximal.
It is interesting to notice that the points of maximum entropy 
$\Gamma=0$ are relatively close to the maxima of the susceptibility. 
%The shaded region of large entropy in Fig.~\ref{fig:gamma} is dominated by 
%the competition between quantum criticality and vortex physics.
However, the maximum entropy region is {\it not}  exactly at the value of the critical field
as is the case for other systems without an ordered phase, as in the Ising chain.\cite{PhysRevB.72.205129}  Nor are those points associated with the finite temperature BKT phase transition
as would be the case for ordered systems in $D>2$.\cite{PhysRevB.72.205129}
The situation in $D=2$ is therefore special, since in this case the
sign change in the cooling rate $\Gamma=0$ signals a maximum in the entropy in the 
crossover region where quantum critical behavior competes with 
vortex excitations in the shaded parameter range in Fig.~\ref{fig:gamma} (bottom).

\begin{figure}[t]
  \centering
  \includegraphics[width=\columnwidth]{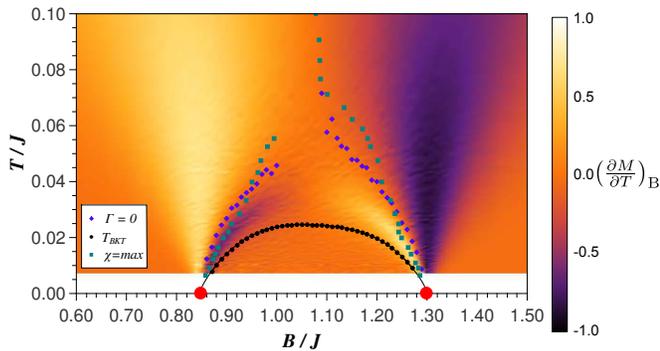}
  \caption[$T$-$B$ phase diagram for the dimerized columnar square lattice]{(Color online) 
The magneto-caloric
derivative $\partial M/\partial T$ in the $T$-$B$ parameter space for $N = 676$.
The BKT transitions $T_{\rm BKT}$ is marked by connected dots (black), points of
 maximum entropy $\Gamma=0$ by diamonds (violet) and maxima in the susceptibility by
squares (green).  } \label{fig:TBphase}
\end{figure}

\section{Conclusions}
In summary, the magneto-caloric quantity  $\partial M/\partial T$ 
turns out to be a universal indicator of the quantum critical behavior.
We plot this quantity in Fig.~\ref{fig:TBphase} in the relevant $T$-$B$ parameter
space. 
On the one hand we have seen in Sec.~\ref{sec:CriticalFields} that the critical scaling is
defined by a linear behavior of $M(T)\propto T$, which leads to a constant and 
large derivative $\partial M/\partial T$.  The regions with quantum critical behavior 
therefore show 
up clearly in Fig.~\ref{fig:TBphase} as the lightest and darkest region in the phase diagram 
above $B_c$ and $B_s$, respectively.
The points of $\Gamma \propto \partial M/\partial T =0$ mark the 
boundaries towards regions, which are dominated by BKT vortex excitations.
These points coincide with the maxima in the susceptibility, but are {\it not}
directly associated with the QPT or the finite temperature BKT phase transition.
The BKT phase transition occurs at significantly lower temperatures and is not
reflected by any directly measurable thermodynamic quantity.\cite{lang}
Nonetheless, the predicted and well established behavior of the spin stiffness 
at the BKT transition holds also for the dimer system, but strong corrections start
to appear at small magnetization (i.e.~boson density) as discussed in Sec.~\ref{sec:BKT}.

We would like to emphasize that magneto-caloric measurements of $\partial M/\partial T$
not only allow a detailed analysis of the QPT, but also are potentially 
a very useful experimental tool
in order 
to identify the effective dimensionality of the underlying spin systems due to the 
different density of states.
In particular, for quasi one-dimensional systems $\partial M/\partial T \propto 1/\sqrt{T}$ 
shows a characteristic divergence above the QPT\cite{} while for $D=3$ we find an 
increase $\partial M/\partial T \propto \sqrt{T}$ analogous 
to the famous $T^{3/2}$ Bloch law.  We find in our numerical simulations 
that $D=2$ is characterized by
perfectly linear behavior above the QPT, i.e.~$\partial M/\partial T=\rm const.$ without any detectable  
logarithmic corrections in contrast to the field theory prediction in Eq.~(\ref{MT}).\cite{PhysRevB.50.258}
As discussed in Sec.~\ref{sec:CriticalFields} this can be used to determine the exact positions of the critical field,  which in turn
allows the quantitative estimate of higher order terms in the analytical
expressions as a function of the antiferromagnetic coupling constants.

\begin{acknowledgments}\label{sec:ackn}
We are thankful for useful discussions with  Achim Rosch, Markus Garst, Denis Morath, 
and Axel Pelster.
This work was supported by the SFB Transregio 49 of the DFG and the 
''Allianz f\"ur Hochleistungsrechnen Rheinland-Pfalz'' (AHRP). 
\end{acknowledgments}


\begin{thebibliography}{27}
\expandafter\ifx\csname natexlab\endcsname\relax\def\natexlab#1{#1}\fi
\expandafter\ifx\csname bibnamefont\endcsname\relax
  \def\bibnamefont#1{#1}\fi
\expandafter\ifx\csname bibfnamefont\endcsname\relax
  \def\bibfnamefont#1{#1}\fi
\expandafter\ifx\csname citenamefont\endcsname\relax
  \def\citenamefont#1{#1}\fi
\expandafter\ifx\csname url\endcsname\relax
  \def\url#1{\texttt{#1}}\fi
\expandafter\ifx\csname urlprefix\endcsname\relax\def\urlprefix{URL }\fi
\providecommand{\bibinfo}[2]{#2}
\providecommand{\eprint}[2][]{\url{#2}}

\bibitem{ott} A. Vogler, {\it et. al}, Phys. Rev. Lett. {\bf 113}, 215301 (2014).

\bibitem{dalibard} Z. Hadzibabic, {\it et. al}, Nature {\bf 441},  1118 (2006).

\bibitem{lang} U. Tutsch, {\it et. al}, Nature Comm. {\bf 5}. 5169 (2014).


\bibitem{sachdev} S. Sachdev, Science {\bf 288}, 475 (2000).

\bibitem[{\citenamefont{et~al.}(2003)}]{Ruegg03p62}
\bibinfo{author}{\bibfnamefont{C.~Ruegg} \bibnamefont{et~al.}},
  \bibinfo{journal}{Nature} \textbf{\bibinfo{volume}{423}}, \bibinfo{pages}{62}
  (\bibinfo{year}{2003}).

\bibitem[{\citenamefont{Wessel et~al.}(2001)\citenamefont{Wessel, Olshanii, and
  Haas}}]{PhysRevLett.87.206407}
\bibinfo{author}{\bibfnamefont{S.}~\bibnamefont{Wessel}},
  \bibinfo{author}{\bibfnamefont{M.}~\bibnamefont{Olshanii}}, \bibnamefont{and}
  \bibinfo{author}{\bibfnamefont{S.}~\bibnamefont{Haas}},
  \bibinfo{journal}{Phys. Rev. Lett.} \textbf{\bibinfo{volume}{87}},
  \bibinfo{pages}{206407} (\bibinfo{year}{2001}).

\bibitem[{\citenamefont{Amaya et~al.}(1969)\citenamefont{Amaya, Tokunaga,
  Yamada, Ajiro, and Haseda}}]{Haseda1969}
\bibinfo{author}{\bibfnamefont{K.}~\bibnamefont{Amaya}},
  \bibinfo{author}{\bibfnamefont{Y.}~\bibnamefont{Tokunaga}},
  \bibinfo{author}{\bibfnamefont{R.}~\bibnamefont{Yamada}},
  \bibinfo{author}{\bibfnamefont{Y.}~\bibnamefont{Ajiro}}, \bibnamefont{and}
  \bibinfo{author}{\bibfnamefont{T.}~\bibnamefont{Haseda}},
  \bibinfo{journal}{Physics Letters A} \textbf{\bibinfo{volume}{28}},
  \bibinfo{pages}{732 } (\bibinfo{year}{1969}), ISSN \bibinfo{issn}{0375-9601}.

\bibitem[{\citenamefont{Tachiki and Yamada}(1970)}]{Tachiki1970}
\bibinfo{author}{\bibfnamefont{M.}~\bibnamefont{Tachiki}} \bibnamefont{and}
  \bibinfo{author}{\bibfnamefont{T.}~\bibnamefont{Yamada}},
  \bibinfo{journal}{Journal of the Physical Society of Japan}
  \textbf{\bibinfo{volume}{28}}, \bibinfo{pages}{1413} (\bibinfo{year}{1970}).

\bibitem[{\citenamefont{Berezinskii}(1971)}]{Bere1}
\bibinfo{author}{\bibfnamefont{V.}~\bibnamefont{Berezinskii}},
  \bibinfo{journal}{Sov. Phys. JETP} \textbf{\bibinfo{volume}{32}},
  \bibinfo{pages}{493} (\bibinfo{year}{1971}).

\bibitem[{\citenamefont{Berezinskii}(1972)}]{Bere2}
\bibinfo{author}{\bibfnamefont{V.}~\bibnamefont{Berezinskii}},
  \bibinfo{journal}{Sov. Phys. JETP} \textbf{\bibinfo{volume}{34}},
  \bibinfo{pages}{610} (\bibinfo{year}{1972}).

\bibitem[{\citenamefont{Kosterlitz and Thouless}(1973)}]{KT1}
\bibinfo{author}{\bibfnamefont{J.~M.} \bibnamefont{Kosterlitz}}
  \bibnamefont{and} \bibinfo{author}{\bibfnamefont{D.~J.}
  \bibnamefont{Thouless}}, \bibinfo{journal}{Journal of Physics C: Solid State
  Physics} \textbf{\bibinfo{volume}{6}}, \bibinfo{pages}{1181}
  (\bibinfo{year}{1973}).

\bibitem[{\citenamefont{Kosterlitz}(1974)}]{KT2}
\bibinfo{author}{\bibfnamefont{J.~M.} \bibnamefont{Kosterlitz}},
  \bibinfo{journal}{Journal of Physics C: Solid State Physics}
  \textbf{\bibinfo{volume}{7}}, \bibinfo{pages}{1046} (\bibinfo{year}{1974}).

\bibitem{troyer} K. Bernardet {\it et al.}. \prb {\bf 65}, 104519 (2002).

\bibitem[{\citenamefont{Derzhko et~al.}(2013)\citenamefont{Derzhko, Richter,
  Krupnitska, and Krokhmalskii}}]{richter}
\bibinfo{author}{\bibfnamefont{O.}~\bibnamefont{Derzhko}},
  \bibinfo{author}{\bibfnamefont{J.}~\bibnamefont{Richter}},
  \bibinfo{author}{\bibfnamefont{O.}~\bibnamefont{Krupnitska}},
  \bibnamefont{and}
  \bibinfo{author}{\bibfnamefont{T.}~\bibnamefont{Krokhmalskii}},
  \bibinfo{journal}{Phys. Rev. B} \textbf{\bibinfo{volume}{88}},
  \bibinfo{pages}{094426} (\bibinfo{year}{2013}).

\bibitem[{\citenamefont{Sachdev}(2011)}]{SachdevQPT}
\bibinfo{author}{\bibfnamefont{S.}~\bibnamefont{Sachdev}},
  \emph{\bibinfo{title}{Quantum Phase Transitions}}
  (\bibinfo{publisher}{Cambridge University Press}, \bibinfo{year}{2011}).


\bibitem[{\citenamefont{Sachdev et~al.}(1994)\citenamefont{Sachdev, Senthil,
  and Shankar}}]{PhysRevB.50.258}
\bibinfo{author}{\bibfnamefont{S.}~\bibnamefont{Sachdev}},
  \bibinfo{author}{\bibfnamefont{T.}~\bibnamefont{Senthil}}, \bibnamefont{and}
  \bibinfo{author}{\bibfnamefont{R.}~\bibnamefont{Shankar}},
  \bibinfo{journal}{Phys. Rev. B} \textbf{\bibinfo{volume}{50}},
  \bibinfo{pages}{258} (\bibinfo{year}{1994}).

\bibitem[{\citenamefont{Popov}(1983)}]{PopovBook}
\bibinfo{author}{\bibfnamefont{V.~N.} \bibnamefont{Popov}},
  \emph{\bibinfo{title}{Functional Integrals in Quantum Field Theory and
  Statistical Physics}} (\bibinfo{publisher}{D. Reidel Publishing Company},
  \bibinfo{year}{1983}).

\bibitem[{\citenamefont{Fisher and Hohenberg}(1988)}]{PhysRevB.37.4936}
\bibinfo{author}{\bibfnamefont{D.~S.} \bibnamefont{Fisher}} \bibnamefont{and}
  \bibinfo{author}{\bibfnamefont{P.~C.} \bibnamefont{Hohenberg}},
  \bibinfo{journal}{Phys. Rev. B} \textbf{\bibinfo{volume}{37}},
  \bibinfo{pages}{4936} (\bibinfo{year}{1988}).

\bibitem[{\citenamefont{Sachdev and Dunkel}(2006)}]{PhysRevB.73.085116}
\bibinfo{author}{\bibfnamefont{S.}~\bibnamefont{Sachdev}} \bibnamefont{and}
  \bibinfo{author}{\bibfnamefont{E.~R.} \bibnamefont{Dunkel}},
  \bibinfo{journal}{Phys. Rev. B} \textbf{\bibinfo{volume}{73}},
  \bibinfo{pages}{085116} (\bibinfo{year}{2006}).

\bibitem[{\citenamefont{Prokof'ev and Svistunov}(2002)}]{PhysRevA.66.043608}
\bibinfo{author}{\bibfnamefont{N.}~\bibnamefont{Prokof'ev}} \bibnamefont{and}
  \bibinfo{author}{\bibfnamefont{B.}~\bibnamefont{Svistunov}},
  \bibinfo{journal}{Phys. Rev. A} \textbf{\bibinfo{volume}{66}},
  \bibinfo{pages}{043608} (\bibinfo{year}{2002}).

\bibitem[{\citenamefont{Prokof'ev et~al.}(2001)\citenamefont{Prokof'ev,
  Ruebenacker, and Svistunov}}]{PhysRevLett.87.270402}
\bibinfo{author}{\bibfnamefont{N.}~\bibnamefont{Prokof'ev}},
  \bibinfo{author}{\bibfnamefont{O.}~\bibnamefont{Ruebenacker}},
  \bibnamefont{and}
  \bibinfo{author}{\bibfnamefont{B.}~\bibnamefont{Svistunov}},
  \bibinfo{journal}{Phys. Rev. Lett.} \textbf{\bibinfo{volume}{87}},
  \bibinfo{pages}{270402} (\bibinfo{year}{2001}).

\bibitem[{\citenamefont{Sachdev and Demler}(2004)}]{PhysRevB.69.144504}
\bibinfo{author}{\bibfnamefont{S.}~\bibnamefont{Sachdev}} \bibnamefont{and}
  \bibinfo{author}{\bibfnamefont{E.}~\bibnamefont{Demler}},
  \bibinfo{journal}{Phys. Rev. B} \textbf{\bibinfo{volume}{69}},
  \bibinfo{pages}{144504} (\bibinfo{year}{2004}).

\bibitem[{\citenamefont{Sachdev}(1999)}]{PhysRevB.59.14054}
\bibinfo{author}{\bibfnamefont{S.}~\bibnamefont{Sachdev}},
  \bibinfo{journal}{Phys. Rev. B} \textbf{\bibinfo{volume}{59}},
  \bibinfo{pages}{14054} (\bibinfo{year}{1999}).

\bibitem[{\citenamefont{Harris}(1973)}]{PhysRevB.7.3166}
\bibinfo{author}{\bibfnamefont{A.~B.} \bibnamefont{Harris}},
  \bibinfo{journal}{Phys. Rev. B} \textbf{\bibinfo{volume}{7}},
  \bibinfo{pages}{3166} (\bibinfo{year}{1973}).

\bibitem[{\citenamefont{Barnes et~al.}(1999)\citenamefont{Barnes, Riera, and
  Tennant}}]{PhysRevB.59.11384}
\bibinfo{author}{\bibfnamefont{T.}~\bibnamefont{Barnes}},
  \bibinfo{author}{\bibfnamefont{J.}~\bibnamefont{Riera}}, \bibnamefont{and}
  \bibinfo{author}{\bibfnamefont{D.~A.} \bibnamefont{Tennant}},
  \bibinfo{journal}{Phys. Rev. B} \textbf{\bibinfo{volume}{59}},
  \bibinfo{pages}{11384} (\bibinfo{year}{1999}).

\bibitem[{\citenamefont{Reigrotzki et~al.}(1994)\citenamefont{Reigrotzki,
  Tsunetsugu, and Rice}}]{Rotzki}
\bibinfo{author}{\bibfnamefont{M.}~\bibnamefont{Reigrotzki}},
  \bibinfo{author}{\bibfnamefont{H.}~\bibnamefont{Tsunetsugu}},
  \bibnamefont{and} \bibinfo{author}{\bibfnamefont{T.~M.} \bibnamefont{Rice}},
  \bibinfo{journal}{Journal of Physics: Condensed Matter}
  \textbf{\bibinfo{volume}{6}}, \bibinfo{pages}{9235} (\bibinfo{year}{1994}).

\bibitem[{\citenamefont{Sylju\aa{}sen and Sandvik}(2002)}]{Sandvik2002}
\bibinfo{author}{\bibfnamefont{O.~F.} \bibnamefont{Sylju\aa{}sen}}
  \bibnamefont{and} \bibinfo{author}{\bibfnamefont{A.~W.}
  \bibnamefont{Sandvik}}, \bibinfo{journal}{Phys. Rev. E}
  \textbf{\bibinfo{volume}{66}}, \bibinfo{pages}{046701}
  (\bibinfo{year}{2002}).

\bibitem[{\citenamefont{Matsumoto and Nishimura}(1998)}]{Mersenne1998}
\bibinfo{author}{\bibfnamefont{M.}~\bibnamefont{Matsumoto}} \bibnamefont{and}
  \bibinfo{author}{\bibfnamefont{T.}~\bibnamefont{Nishimura}},
  \bibinfo{journal}{ACM Trans. Model. Comput. Simul.}
  \textbf{\bibinfo{volume}{8}}, \bibinfo{pages}{3} (\bibinfo{year}{1998}), ISSN
  \bibinfo{issn}{1049-3301}.

\bibitem[{\citenamefont{Pollock and Ceperley}(1987)}]{PhysRevB.36.8343}
\bibinfo{author}{\bibfnamefont{E.~L.} \bibnamefont{Pollock}} \bibnamefont{and}
  \bibinfo{author}{\bibfnamefont{D.~M.} \bibnamefont{Ceperley}},
  \bibinfo{journal}{Phys. Rev. B} \textbf{\bibinfo{volume}{36}},
  \bibinfo{pages}{8343} (\bibinfo{year}{1987}).

\bibitem[{\citenamefont{Sandvik}(1997)}]{PhysRevB.56.11678}
\bibinfo{author}{\bibfnamefont{A.~W.} \bibnamefont{Sandvik}},
  \bibinfo{journal}{Phys. Rev. B} \textbf{\bibinfo{volume}{56}},
  \bibinfo{pages}{11678} (\bibinfo{year}{1997}).

\bibitem{helicity} M.E. Fisher, M.N. Barber, and D. Jasnow, \pra {\bf 8}, 1111 (1973).

\bibitem[{\citenamefont{Prokof'ev and Svistunov}(2000)}]{PhysRevB.61.11282}
\bibinfo{author}{\bibfnamefont{N.~V.} \bibnamefont{Prokof'ev}}
  \bibnamefont{and} \bibinfo{author}{\bibfnamefont{B.~V.}
  \bibnamefont{Svistunov}}, \bibinfo{journal}{Phys. Rev. B}
  \textbf{\bibinfo{volume}{61}}, \bibinfo{pages}{11282} (\bibinfo{year}{2000}).

\bibitem[{\citenamefont{Melko et~al.}(2004)\citenamefont{Melko, Sandvik, and
  Scalapino}}]{PhysRevB.69.014509}
\bibinfo{author}{\bibfnamefont{R.~G.} \bibnamefont{Melko}},
  \bibinfo{author}{\bibfnamefont{A.~W.} \bibnamefont{Sandvik}},
  \bibnamefont{and} \bibinfo{author}{\bibfnamefont{D.~J.}
  \bibnamefont{Scalapino}}, \bibinfo{journal}{Phys. Rev. B}
  \textbf{\bibinfo{volume}{69}}, \bibinfo{pages}{014509}
  (\bibinfo{year}{2004}).

\bibitem[{\citenamefont{Weber and Minnhagen}(1988)}]{PhysRevB.37.5986}
\bibinfo{author}{\bibfnamefont{H.}~\bibnamefont{Weber}} \bibnamefont{and}
  \bibinfo{author}{\bibfnamefont{P.}~\bibnamefont{Minnhagen}},
  \bibinfo{journal}{Phys. Rev. B} \textbf{\bibinfo{volume}{37}},
  \bibinfo{pages}{5986} (\bibinfo{year}{1988}).

\bibitem[{\citenamefont{Harada and Kawashima}(1998)}]{JPSJ.67.2768}
\bibinfo{author}{\bibfnamefont{K.}~\bibnamefont{Harada}} \bibnamefont{and}
  \bibinfo{author}{\bibfnamefont{N.}~\bibnamefont{Kawashima}},
  \bibinfo{journal}{J. Phys. Soc. Jap.}
  \textbf{\bibinfo{volume}{67}}, \bibinfo{pages}{2768} (\bibinfo{year}{1998}).

\bibitem[{\citenamefont{Cuccoli
  et~al.}(2003{\natexlab{b}})\citenamefont{Cuccoli, Roscilde, Tognetti, Vaia,
  and Verrucchi}}]{PhysRevB.67.104414}
\bibinfo{author}{\bibfnamefont{A.}~\bibnamefont{Cuccoli}},
  \bibinfo{author}{\bibfnamefont{T.}~\bibnamefont{Roscilde}},
  \bibinfo{author}{\bibfnamefont{V.}~\bibnamefont{Tognetti}},
  \bibinfo{author}{\bibfnamefont{R.}~\bibnamefont{Vaia}}, \bibnamefont{and}
  \bibinfo{author}{\bibfnamefont{P.}~\bibnamefont{Verrucchi}},
  \bibinfo{journal}{Phys. Rev. B} \textbf{\bibinfo{volume}{67}},
  \bibinfo{pages}{104414} (\bibinfo{year}{2003}{\natexlab{b}}).

\bibitem{bkt-sandvik} Y.-D. Hsieh, Y.-J. Kao, A.W. Sandvik, J. Stat. Mech. (2013) P09001.

\bibitem[{\citenamefont{Warburg}(1881)}]{Warburg1881}
\bibinfo{author}{\bibfnamefont{E.}~\bibnamefont{Warburg}}, \bibinfo{journal}{J.
  Phys. Theor. Appl.} \textbf{\bibinfo{volume}{10}} (\bibinfo{year}{1881}).


\bibitem[{\citenamefont{Garst and Rosch}(2005)}]{PhysRevB.72.205129}
\bibinfo{author}{\bibfnamefont{M.}~\bibnamefont{Garst}} \bibnamefont{and}
  \bibinfo{author}{\bibfnamefont{A.}~\bibnamefont{Rosch}},
  \bibinfo{journal}{Phys. Rev. B} \textbf{\bibinfo{volume}{72}},
  \bibinfo{pages}{205129} (\bibinfo{year}{2005}).

\end{thebibliography}
\end{document}